\documentclass[conference]{IEEEtran}
\usepackage{cite}
\usepackage{amsmath,amssymb,amsfonts}
\usepackage{algorithmic}
\usepackage{graphicx}
\usepackage{textcomp}
\usepackage{color,soul}
\usepackage{subfig}
\usepackage{multirow}

\def\BibTeX{{\rm B\kern-.05em{\sc i\kern-.025em b}\kern-.08em
    T\kern-.1667em\lower.7ex\hbox{E}\kern-.125emX}}

\newcommand{\comment}[1]{}

\newcommand{\figvs}[4]{\begin{figure}[!t]
\centering
\includegraphics[width=#1\columnwidth,keepaspectratio,#3]{#2.pdf}
\caption{#4}
\label{#2}
\end{figure}}

\newcommand{\figfull}[4]{\begin{figure*}[!t]
\centering
\includegraphics[width=#1\columnwidth,keepaspectratio,#3]{#2.pdf}
\caption{#4}
\label{#2}
\end{figure*}}

\begin{document}

\title{A Case for High Performance Overlays: Deep Learning and the Power of FPGAs}
\title{DLA: Compiler and FPGA Overlay for Neural Network Inference Acceleration}

\author{
\IEEEauthorblockN{Mohamed S. Abdelfattah, David Han, Andrew Bitar, Roberto DiCecco, Shane O'Connell,\\
Nitika Shanker, Joseph Chu, Ian Prins, Joshua Fender, Andrew C. Ling, Gordon R. Chiu}
\IEEEauthorblockA{
\textit{Programmable Solutions Group, Intel} \\
Toronto, Canada \\
\{firstname.lastname\}@intel.com}
}

\maketitle

\begin{abstract}
Overlays have shown significant promise for field-programmable gate-arrays (FPGAs) as they allow for fast development cycles and remove many of the challenges of the traditional FPGA hardware design flow.
However, this often comes with a significant performance burden resulting in very little adoption of overlays for practical applications.
In this paper, we tailor an overlay to a specific application domain, and we show how we maintain its full programmability without paying for the performance overhead traditionally associated with overlays.
Specifically, we introduce an overlay targeted for deep neural network inference with only \texttildelow1\% overhead to support the control and reprogramming logic using a lightweight very-long instruction word (VLIW) network.
Additionally, we implement a sophisticated domain specific graph compiler that compiles deep learning languages such as Caffe or Tensorflow to easily target our overlay.
We show how our graph compiler performs architecture-driven software optimizations to significantly boost performance of both convolutional and recurrent neural networks (CNNs/RNNs) -- we demonstrate a 3$\times$ improvement on ResNet-101 and a 12$\times$ improvement for long short-term memory (LSTM) cells, compared to na\"ive implementations.
Finally, we describe how we can tailor our hardware overlay, and use our graph compiler to achieve \texttildelow900~fps on GoogLeNet on an Intel Arria~10 1150 -- the fastest ever reported on comparable FPGAs.
\end{abstract}

\section{Introduction}
\figfull{1.9}{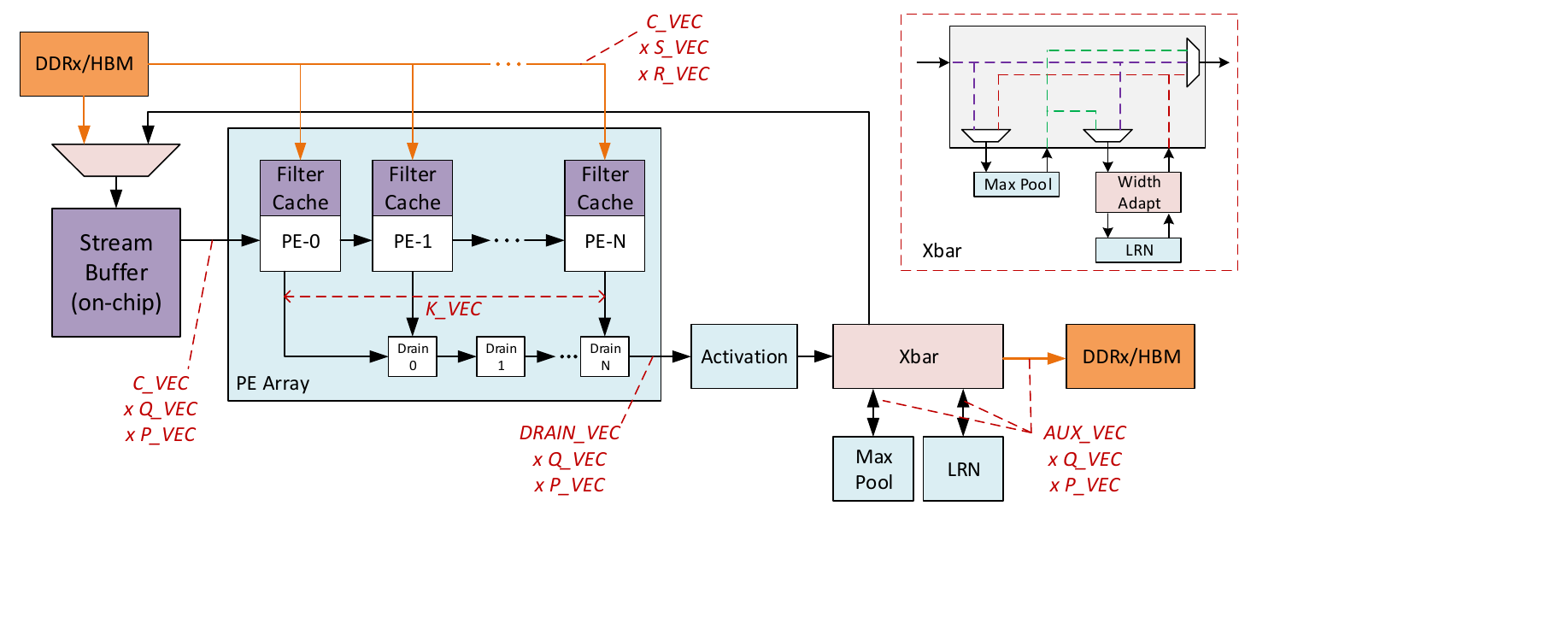}{trim = 0cm 2.5cm 5cm 0cm}{System-level diagram of our neural network inference accelerator (DLA).}

Creating custom high-performance hardware designs on field-programmable gate arrays (FPGAs) is difficult and time-consuming when compared to software-programmable devices such as CPUs.
A hardware designer must describe their system in a cycle-accurate manner, and worry about low-level hardware considerations such as timing closure to memory interfaces.
Over the past decade, significant progress has been made in easing the use of FPGAs through high-level languages such as OpenCL, making it easier to implement high-performance designs~\cite{opencl}.
However, even when using high-level design, one must still carefully describe an efficient \textit{parallel} hardware architecture that leverages the FPGA's capabilities such as the massive on-chip memory bandwidth or configurable multiplier blocks.
Additionally, the designer must optimize both area and frequency through long compilations to realize performance gains versus other programmable platforms.
Compared to writing a software algorithm targeting a CPU, designing for FPGAs is still drastically more difficult.
Our goal in this paper is to present a software-programmable hardware overlay on FPGAs to realize the ease-of-use of software programmability and the efficiency of custom hardware design.

\comment{
Overlays have shown significant promise as a means to empower FPGA users with a software development flow that comes with fast development cycles and removes the need to solve the most challenging problems with FPGA design such as timing closure~\cite{timing_closure}.
However, recent work has shown that this comes with a significant performance burden and often leads to designs that are 50-80\%~\cite{bad_overlay} slower when compared against designs implemented directly on the FPGA.
As a result, there has been little adoption of overlays in industry and traditional hardware design flows still dominate FPGA use.
}

We introduce a domain specific approach to overlays that leverages both software and hardware optimizations to achieve state-of-the-art performance on the FPGA for neural network (NN) acceleration.
For hardware, we partition configurable parameters into runtime and compile time parameters such that you can tune the architecture for performance at compile time, and program the overlay at runtime to accelerate different NNs.
We do this through a lightweight very-long instruction word (VLIW) network that delivers full reprogrammability to our overlay without incurring \textit{any} performance or efficiency overhead (typical overlays have large overhead~\cite{bad_overlay}).
Additionally, we create a flexible architecture where only the core functions required by a NN are connected to a parameterizable interconnect (called Xbar).
This avoids the need to include all possible functions in our overlay during runtime; rather, we can pick from our library of optimized kernels based on the group of NNs that are going to run on our system.
Our approach is unlike previous work that created hardware that can only run a single/specific NN~\cite{previous_work, cnn1, cnn2}.

On the software side, we introduce an architecture-aware graph compiler that efficiently maps a NN to the overlay.
This both maximizes the hardware efficiency when running the design and simplifies the usability of the end application, where users are only required to enter domain specific deep learning languages, such as Caffe or Tensorflow, to program the overlay.
Our compiler generates VLIW instructions that are loaded into the FPGA and used for reprogramming the overlay in tens of clock cycles thus incurring no performance overhead.
Compared to fixed-function accelerators that can only execute one NN per application run, our approach opens the door to allow for multiple NN’s be run consecutively in a single application run~\cite{ssd} by simply reprogramming our overlay instead of recompiling or reconfiguring the FPGA.


The rest of this paper is organized as follows.
Section~\ref{sec_hw} introduces our hardware architecture.
We describe how we target specific NNs using our compile-time parameters and Xbar interconnect.
Importantly, we describe our lightweight VLIW network in Section~\ref{sec_config}, used for programming the overlay.
Next, we describe our NN graph compiler in Section~\ref{sec_sw}, and detail some of our architecture-driven optimizations that allow the efficient implementation of NNs on architecture variants of different sizes.
Sections~\ref{sec_resnet} and \ref{sec_lstm} detail how our graph compiler and hardware overlay work together for efficient implementation of CNNs and RNNs.
We walk through hardware and software optimizations in implementing both the ResNet and GoogLeNet CNNs, allowing us to achieve record-setting performance on GoogLeNet.
Finally, we discuss the implementation of a long short-term memory (LSTM) cell by simply adding an additional kernel to our overlay, and relying on our graph compiler to mutate the LSTM cell graph to fit within our overlay.
In this paper, we refer to our system as ``DLA" -- our Deep Learning Accelerator.


\section{Hardware Architecture}
\label{sec_hw}

Our domain specific overlay aims to be general enough to implement any NN, but still remain customizable so that it can be optimized for a specific NN only.
Fig.~\ref{dla_system_diagram} shows an overview of our overlay.
At the core of our overlay is a 1D systolic processing element (PE) array that performs dot product operations in each PE to implement general matrix math such as convolutions or multiplications.
We omit the discussion of numerics in this paper but we support different floating-point formats such as FP32/16/11/10/9/8 which have been shown to work well with inference~\cite{microsoft_brainwave} -- these could be easily modified to support any nascent innovations in data type precisions such as bfloat~\cite{tensorflow}, and other unique fixed or floating point representations, due to the flexible FPGA fabric.

As Fig~\ref{dla_system_diagram} shows, our Xbar interconnect can augment the functionality of our overlay with different \textit{auxiliary} functions (also referred to as \textit{kernels} in this paper).
This section goes through different parts of our hardware architecture and highlights the built-in compile-time flexibility and run-time programmability of our overlay.

\comment{
This section starts by explaining our Xbar interconnect and how it enables easy function extensibility.
We then show how our lightweight VLIW network is used for reprogramming the overlay -- this provides context for the following Section~\ref{sec_sw} where we describe our software NN compiler that generates the VLIW instructions.
We elaborate different aspects of our hardware overlay architecture and show how can change the degree of parallelism or numerical precision to enable parameterizing our overlay to target different FPGAs or NN workloads.
Finally, we describe how we leverage on-chip memory resources to efficiently cache NN inputs and filters for high-performance.
}

\subsection{VLIW Network}
\label{sec_config}
\comment{
\hl{Mohamed Abdelfattah}

\begin{itemize}
  \item Describe configuration ring.
  \item Double-buffering of config data in fifos.
  \item Draw analogy to VLIW, but distributed across kernels.
  \item Lightweight: area overhead is small -- measure.
\end{itemize}
}

To implement a NN on DLA, our graph compiler breaks it into units called ``subgraphs" that fit within the overlay's buffers and compute elements.
For example, with convolutional neural networks (CNNs), a subgraph is typically a single convolution with an optional pooling layer afterwards.
We deliver new VLIW instructions for each subgraph to program DLA correctly for the subgraph execution.

\figvs{0.9}{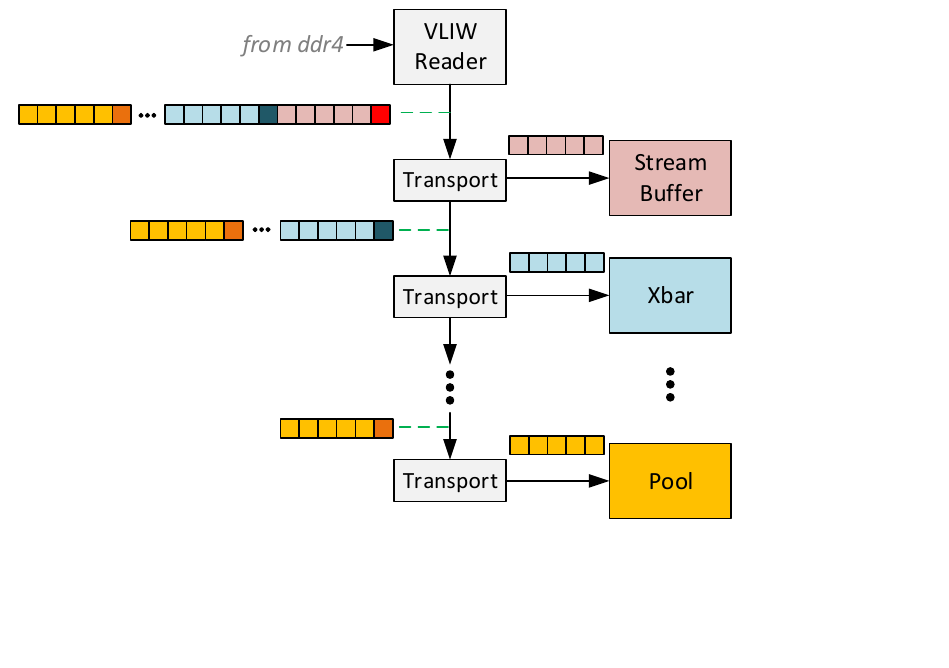}{trim = 0cm 2.5cm 3cm 0cm}{VLIW network distributes instructions to each kernel.}

Our novel VLIW network distributes instructions to each kernel as shown in Fig.~\ref{config_network}.
The VLIW reader continuously fetches the instructions for the next subgraph from external memory and sends it down an 8-bit unidirectional ring network that is connected to all of the kernels in DLA.
The VLIW instruction sequence is divided into different portions for each kernel.
A special header packet identifies the kernel, then it is followed by a series of programming instructions that are destined for that kernel.
The ``Transport" kernels parse the header packet and redirects the instructions that follow to the correct kernel as shown in Fig.~\ref{config_network}.
The transport kernels also assemble the 8-bit packets into 32-wide instructions for direct kernel consumption.

Our \textit{instructions} are actually counter end values and control flags that are directly loaded into registers within each kernel to govern its operation -- this avoids the need for any instruction decode units.
For example, the pool kernel recieves approximately a dozen instructions: the image height/width/depth, the pool window size, and the type of pooling (maxpool or average pool).
Before executing each subgraph, the pool kernel would read each of its 12 instructions serially, consuming 12 clock cycles -- this has no material impact on performance that typically takes thousands of cycles.
However, it ensures that the entire VLIW network can remain only 8 bits wide, with a minimal area overhead of only \texttildelow3000 LUTs -- about 1\% of an Arria-10 1150 FPGA device as shown in Table~\ref{vliw_area}.
Adding new auxiliary programmable functions (kernels) to DLA is simple and has little overhead -- we extend the VLIW network with an additional transport kernel, and connect that new kernel to the Xbar without affecting existing kernels or instructions.

\begin{table}[ht]
\centering
\caption{Area overhead of VLIW network for DLA with 10 kernels at frequency of 450 MHz on Arria 10.}
\label{vliw_area}
\begin{tabular}{|c|r|r|r|}
\hline
 & LUTs & FFs & ALMs\\
\hline
VLIW Reader & 1832 & 1841 & 1473\\
\hline
Transport & 126 & 139 & 73 \\
\hline
Total & 3092 & 3231 & 2046\\
\hline
\end{tabular}
\end{table}


\subsection{Xbar Interconnect}
\label{sec_xbar}

Machine learning is a fast-developing field -- we are increasingly seeing new functions implemented by the machine learning research community.
For example, new activation functions are constantly being evaluated such as ``Swish"~\cite{swish}.
A quick look at Tensorflow shows that there more than 100 different layer types that users can experiment with in building different NNs~\cite{tensorflow}.
We aim to use the Xbar for extensibility of DLA such that users can easily add or remove functions to implement different types of NNs.

Fig.~\ref{dla_system_diagram} shows an example Xbar interconnect used to connect pool/LRN kernels for CNNs.
As the diagram shows, the Xbar is actually a custom interconnect built around exactly what is needed to connect the auxiliary kernels.
For example, the SqueezeNet graph has no local response normalization (LRN) layers, so we can remove that kernel completely.
From a prototxt architecture description, the Xbar (including width adaptation) is automatically created to connect auxiliary kernels.
We use width adapters to control the throughput of each auxiliary kernel -- for example, we can decrease the width of infrequent kernels such as LRN to conserve logic resources.
The interconnection pattern within the Xbar is also customizable based on the order of the auxiliary operations.
For example, the AlexNet graph has both MaxPool and LRN layers, but LRN always comes first; whereas the GoogLeNet graph has some layers in which MaxPool precedes LRN, which is supported by adding more multiplexing logic.

To demonstrate the power of our extensible architecture (and compiler which is presented in Section~\ref{sec_sw}), we add a single kernel to the Xbar in Section~\ref{sec_lstm} which extends our architecture to also implement LSTM cells alongside CNNs -- this allows implementing video-based RNNs commonly used for gesture recognition for instance~\cite{gesture}.



\subsection{Vectorization}

\comment{
\hl{Author: Mohamed}

\begin{itemize}
  \item Crossbar to connect different auxiliary kernels.
  \item Show area and speed of crossbar as we vary the number of things connected to it. (can't run that easily..)
  \item Winograd on or off. You want to turn it off for 1x1 filters -- can also have a dynamic runtime switch for it.
  \item Show the effect of customizing CVEC/PVEC/KVEC etc. The tradeoff is different for different graphs.
\end{itemize}
}


To ensure our overlay can be customized to different neural network models and FPGA devices, we support \textit{vectorization}, or degree of parallelism, across different axes.
Figure~\ref{dla_system_diagram} shows some of the degrees of parallelism available in the accelerator, configurable via vectorization.
Q\_VEC and P\_VEC refer to the parallelism in the width and height dimensions, while C\_VEC and K\_VEC refer to the input/output depth parallelism respectively.
Every clock cycle, we process the product of \{Q\_VEC, P\_VEC, C\_VEC, and K\_VEC\} feature values in parallel.

\figvs{0.95}{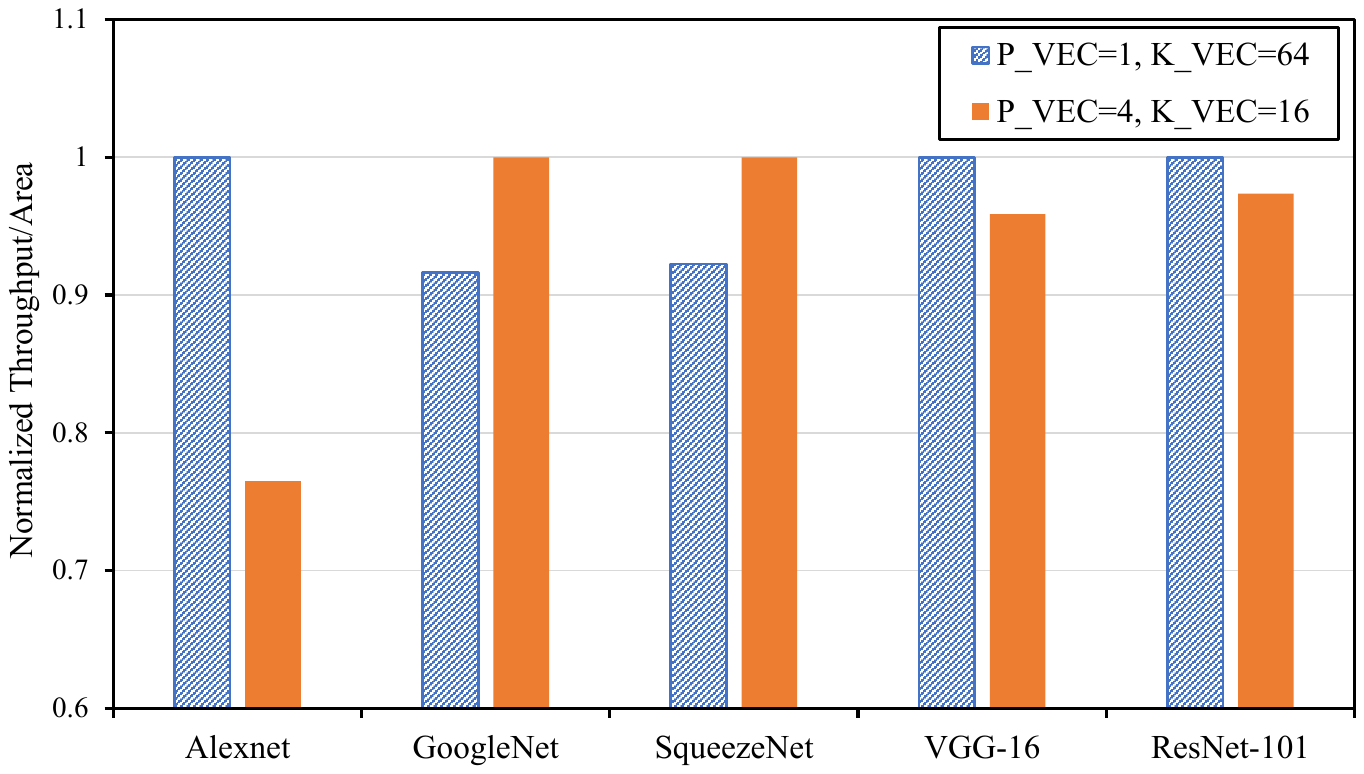}{trim = 1cm 2.3cm 1cm 6.2cm,clip}{Throughput/Area on two architectures with different P\_VEC and K\_VEC vectorization.}

Initially, our design was scaled by increasing K\_VEC; however, this method of scaling saw diminishing returns, since quantization inefficiencies can become more pronounced as vectorization dimensions increase.
For example, if the output depth (K) of a layer is 96, and K\_VEC is 64, this will require 2 complete iterations through the PE array, with only 96/128 (75\%) useful computations.
On the other hand, if K\_VEC is 32, the output depth divides perfectly into 3 iterations at 100\% efficiency.
To mitigate this quantization effect, it is possible to balance the scaling of the design across multiple different dimensions besides just K\_VEC (e.g. P\_VEC, Q\_VEC, C\_VEC, etc).
The optimal balance of vectorization depends on the graph's layer dimensions.
Figure~\ref{kvec} demonstrates this point by comparing the throughput of two architectures with similar area for different graphs.
As the figure shows, the optimal balance of scaling the design between P\_VEC and K\_VEC varies based on the neural network topology being used.
This is an example of how we tune our overlay to get top performance on specific NNs.


\comment{
\subsection{Minifloat Arithmetic}
\input{minifloat}
}

\subsection{Stream Buffer and Filter Caches}


\figvs{1}{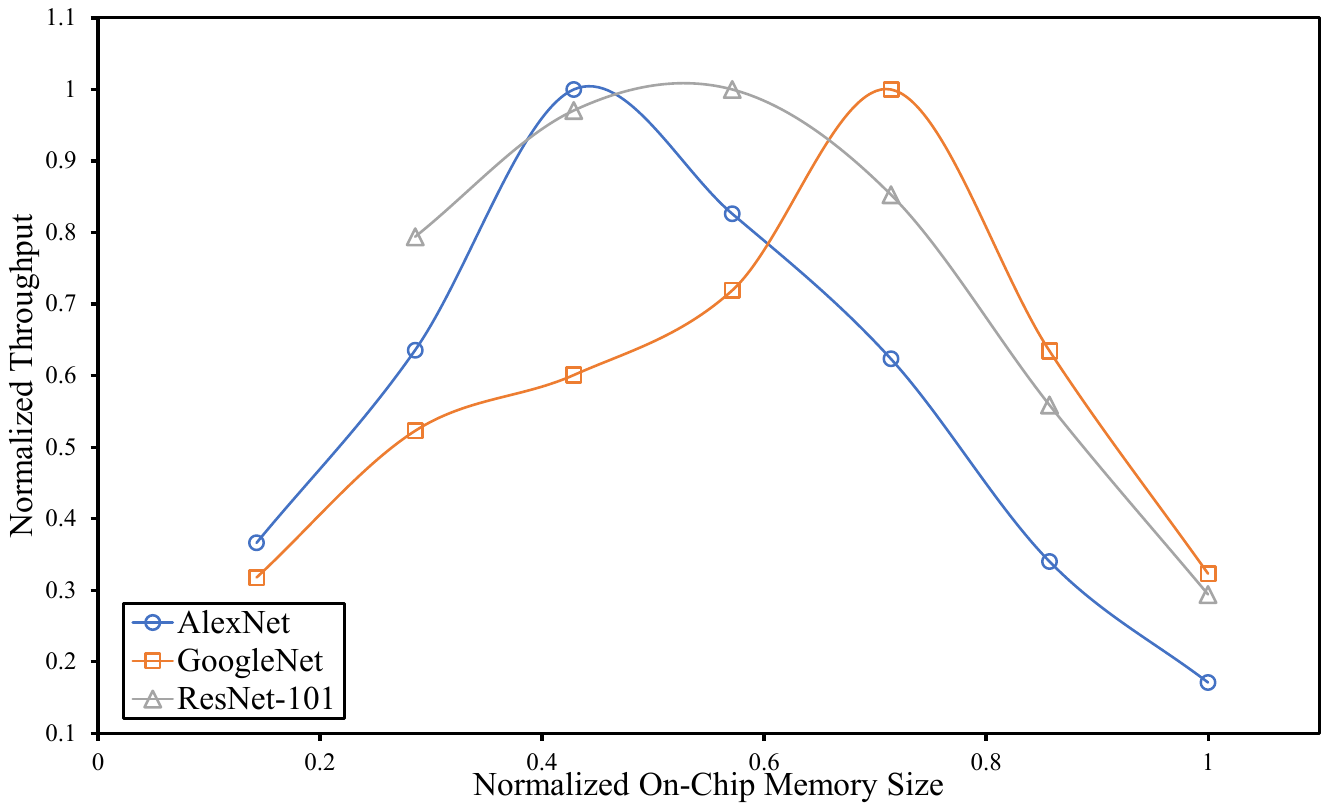}{trim = 1cm 4.0cm 1cm 4.5cm}{Impact of stream buffer memory vs. compute tradeoff on AlexNet, GoogleNet and ResNet-101.}


A single Arria~10 FPGA contains \texttildelow4~TB/s on-chip memory bandwidth, interspersed within the FPGA in configurable 20~Kbit memory blocks.
This powerful FPGA resource is pivotal in determining the performance of FPGA compute operations -- DLA leverages these block RAMs to buffer both activation and filter tensors.
As Fig.~\ref{dla_system_diagram} shows, filters are stored in a double-buffered ``filter cache" contained in each PE, allowing the PEs to compute data while filters are pre-loaded from external memory for the next subgraph.
The ``stream buffer'' is a flexible scratchpad that is used to store intermediate tensors on-chip.
Many of our graph compiler passes are dedicated for efficient use of this stream buffer as Section~\ref{sec_sw} will show.

When presented with an intermediate tensor larger than the stream buffer or filter caches, our graph compiler \textit{slices} the tensor into multiple pieces that fit within our on-chip caches, and the rest of the pieces are stored in slower off-chip memory, and require higher latency to fetch and compute.
To limit this \textit{slicing}, we can increase the size of the stream buffer and/or filter caches, but this decreases the number of RAM blocks available to increase PE array vectorization.
Therefore, there is a memory-vs-compute tradeoff for each NN to balance the size of the caches and the number of PEs -- Fig.~\ref{memory-size-vs-throughput} illustrates this tradeoff for different NNs.
As the figure shows, a tradeoff that is optimal for one NN can cause 40\% or more performance degradation for a second NN.




\section{Graph Compiler}
\label{sec_sw}


The previous section focused on the hardware overlay architecture and how to configure it at compile time to maximize performance for a specific NN graph.
This section describes our NN graph compiler that takes advantage of the overlay VLIW instructions to decompose, optimize, and run a NN model on the overlay.
The graph compiler breaks down a NN into subgraphs, schedules subgraph execution, and importantly, allocates explicit cache buffers to optimize the use of our stream buffer and filter caches.
This section goes through our core compiler ``passes" (slicing, scheduling and allocation), and shows examples of how smart graph compilation allows more efficient hardware implementations.
Besides these general core passes, our compiler implements more specific algorithms that target and optimize specific NN patterns as we show in the following Sections~\ref{sec_resnet} and \ref{sec_lstm}.


\comment{

A DLA hardware overlay is a VLIW accelerator where each data movement and processing step is statically scheduled.
The creation of this schedule is the responsibility of an offline tool called the DLA Graph Compiler.
This tool accepts data flow graph definitions, such as from Tensorflow or Caffe, decomposes the high level instructions into ones executable by single DLA VLIW instructions, schedules instruction execution, allocates explicit cache buffers and schedules the movement of data required by those instructions.
The end goal is to maximize throughput while minimizing latency for a given DLA architecture.

In the rest of this section we describe how large tensor operations are split into sets of smaller ones and illustrate how the schedule and caching allocation scheme impacts both on on-chip cache size requirements as well as performance.

These slicing, scheduling and allocation ``passes" are in the core of our graph compiler, however, our compiler is easily extensible with additional passes that target more specific NN optimizations as we will show in the following sections.
}


\subsection{Slicing}
\label{sec_slicing}

To achieve the highest possible throughput for a given DLA architecture it is desirable to size the stream buffer and filter caches in such a way to fit the entire input feature tensor and filter tensor.
However, as the resolution of images increases and graph topologies for NNs become deeper, on-chip allocation for these tensors may not be feasible.
To overcome this constraint, slices of the input tensor are fetched from external memory into the stream buffer and processed independently by DLA.

The 3D input feature tensor can be sliced along the height, width, or depth to fit in the on-chip stream buffer.
When slicing along the width and height, the slices must overlap if the filter window size is greater than 1x1.
The graph compiler tries to pick slices that minimize the overlapped computation for the sliced tensor.
Alternatively, slicing across the depth does not require overlapped computations, but requires an additive operation to add the results of the depth-wise slices.

\comment{
The 3D input feature tensor can be sliced along the height, width, or depth to fit in the stream buffer.
When slicing along the height or the width, a given slice may incur an additional penalty when fetching data from DDR4 compared to an unsliced input as the input slices may need to overlap due to the windowing operation of convolutions.
DLA slices the input minimally to satisfy on-chip constraints to minimize the cost of the resulting overlap region.
Likewise, when the filter data corresponding to a single filter cannot be allocated in the filter cache we can slice the filter similarly.
However, since the height and width of CNN filters is significantly smaller than that of the input feature tensor, slicing along the depth is preferable.
When we slice along the depth, there is no overlapping region, but an element-wise addition operation is required to combine the partial results of each depth-wise slice.
}

\figvs{0.75}{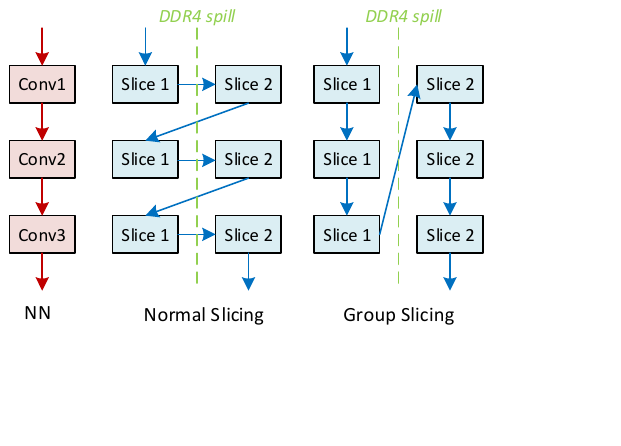}{trim = 0cm 1.8cm 2cm 0cm}{Group slicing minimizes external memory spillpoints by computing multiple sequential convolutions for each slice.}

To boost performance and minimize the number of DDR4 spillpoints, we enhance our slicing algorithm to slice multiple sequential convolutions together (called ``Group Slicing").
Instead of completing all slices within a layer, we compute several sequential convolutions with a single slice using the stream buffer before moving onto the next slice.
Fig.~\ref{group_slicing} illustrates how group slicing reduces the number of external-memory spillpoints for a sample NN.
For Resnet101 with image resolution of 1080p (HD), our Group Slicing algorithm improves throughput by 19\% compared to simple slicing.


\subsection{Allocation}

\figvs{0.9}{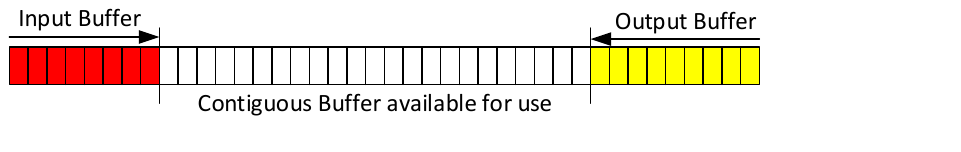}{trim = 0cm 0.5cm 3cm 0cm}{Double-buffering in the stream buffer.}

\begin{figure*}[ht]
  \centering
  \subfloat[Inception module with relative output sizes for each subgraph.]{
  \includegraphics[width=0.2\linewidth]{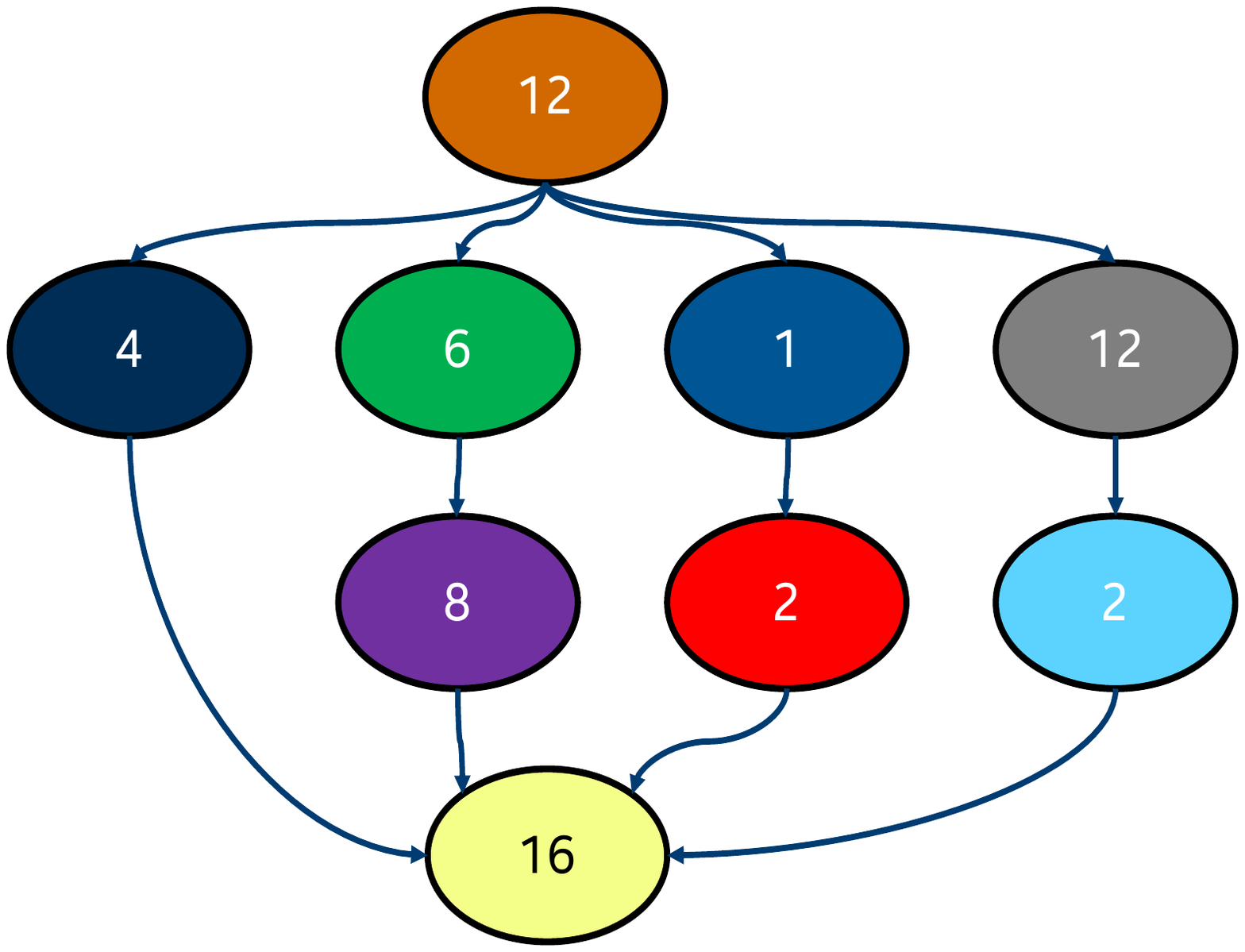}
  \label{fig_inception}
  }
  \subfloat[Stream buffer usage when using a depth first schedule.]{
  \includegraphics[width=0.4\linewidth]{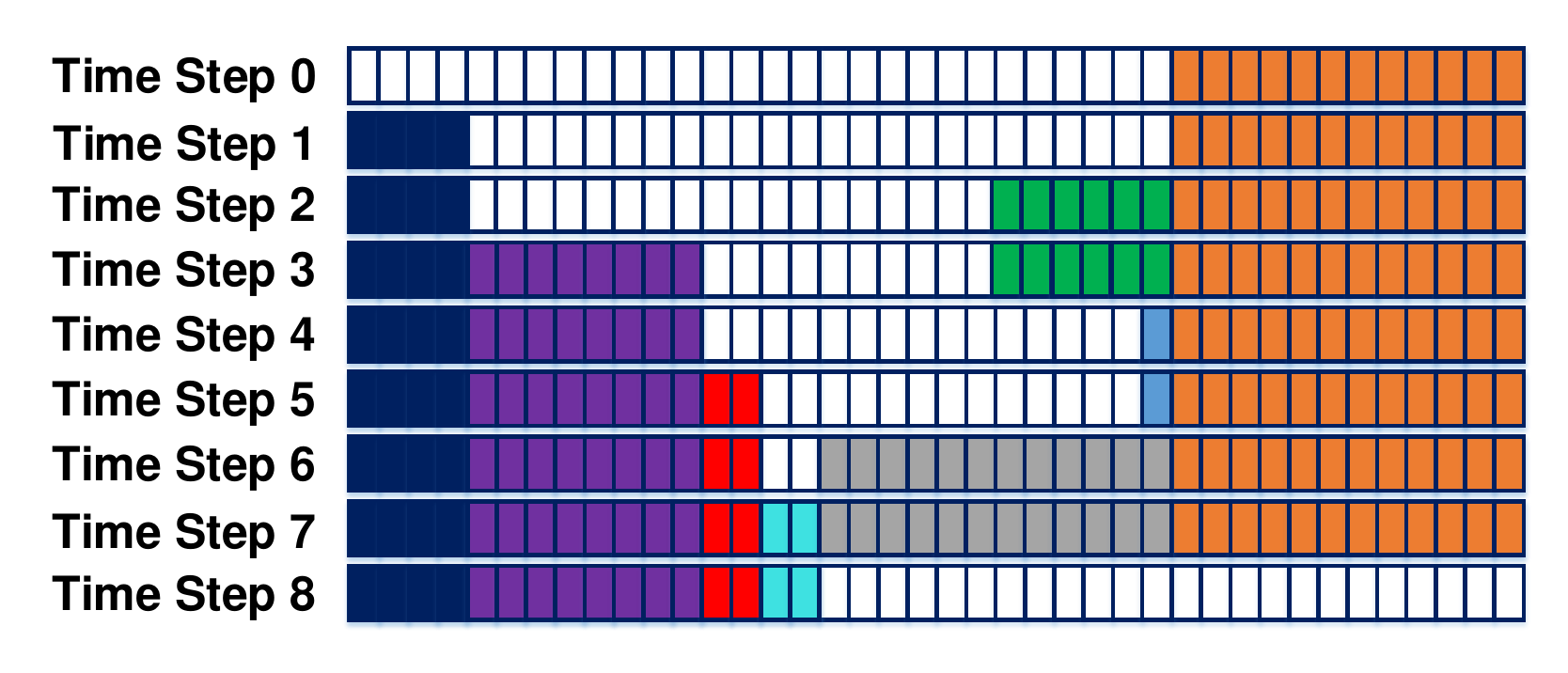}
  \label{fig_dfs_schedule}
  }
  \subfloat[Stream buffer usage with an improved schedule.]{
  \includegraphics[width=0.4\linewidth]{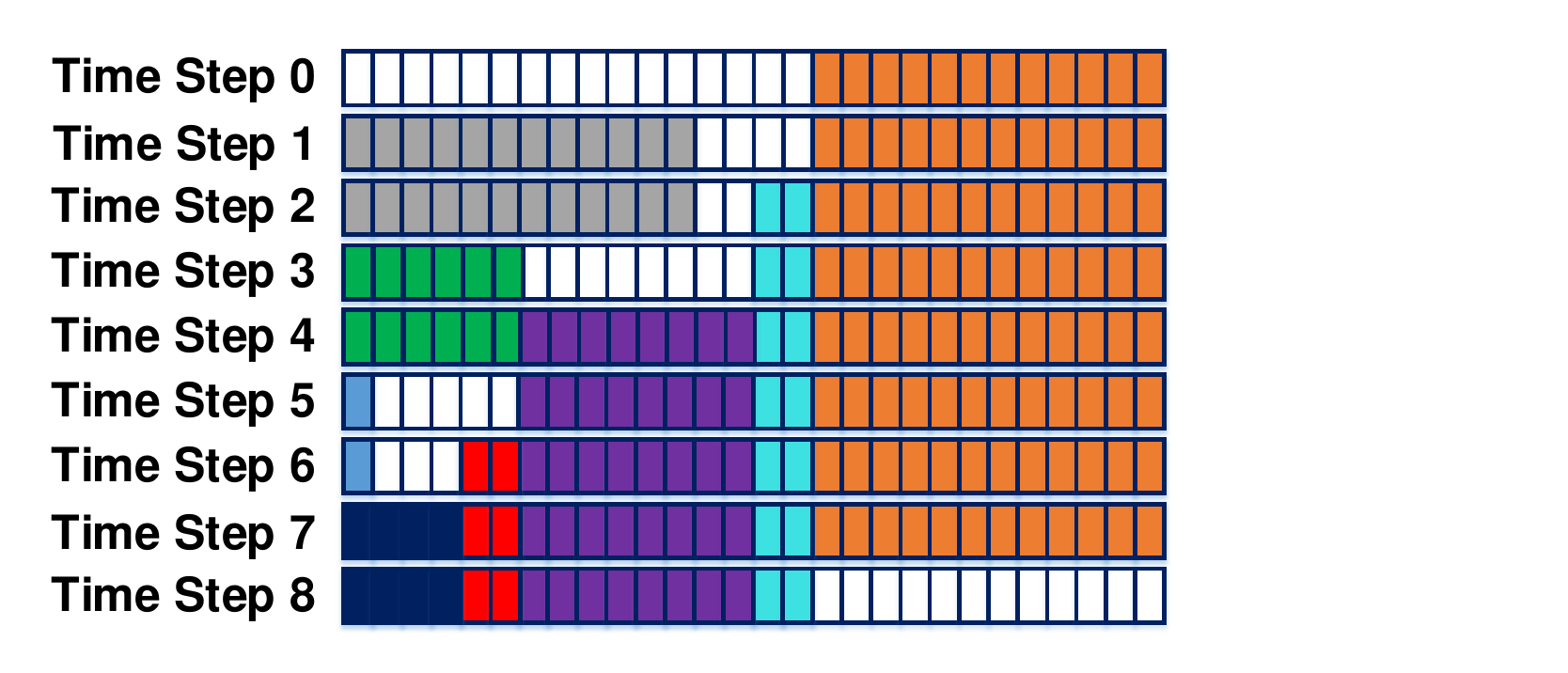}
  \label{fig_improved_schedule}
  }
  \caption{Scheduling one of the GoogLeNet~\cite{szegedy2015googlenet} inception modules.}
  \label{fig_sched}
\end{figure*}

The allocation pass manages reading and writing from the stream buffer.
Allocation calculates the read and write addresses for each slice, and computes the total stream buffer memory used by a graph.
One of the main goals is to reduce fragmentation -- gaps between allocated memory blocks in the stream buffer.
In its most simple operation, the stream buffer is used as a double buffer to store both the input and output of a subgraph.
To achieve this double-buffering while reducing fragmentation, the input buffer starts at address 0 and counts up, while the output buffer starts at the end of the stream buffer and counts down.
As Fig.~\ref{alloc} shows, this leaves a contiguous space in the middle of the stream buffer that can be used to allocate more data slices in the stream buffer; this is especially useful for graphs that have multiple branches as demonstrated by the GoogLeNet example in Section~\ref{sec_sched}.
Note that the allocation pass must keep track of the \textit{lifetime} of each buffer to be able to free/overwrite its memory in the stream buffer once it is no longer used.
Additionally, our allocation pass also assigns addresses in external memory when the stream buffer isn't large enough, but external memory size is not a problem so it is simply done left-to-right, in the first available space.


\subsection{Scheduling}
\label{sec_sched}

The DLA compiler partitions NNs into subgraphs where a subgraph is a list of functions that can be chained together and implemented on DLA without writing to a buffer, except at the very end of the subgraph execution -- scheduling decides when each subgraph is executed.
In the case of early CNN models such as AlexNet~\cite{krizhevsky2012alexnet} or VGG16~\cite{simonyan2014very} there is very little need for a scheduler as there are no decisions to be made on which subgraph to execute next.
When considering CNNs with branching nodes such as GoogLeNet~\cite{szegedy2015googlenet}, ResNet~\cite{he-corr15}, or graphs that require slicing, the order of subgraph execution heavily influences the stream buffer size that is required for a given graph to avoid external memory spill points.

Fig.~\ref{fig_inception} illustrates an example of an inception module from Googlenet, partitioned into DLA subgraphs with the relative output sizes of each subgraph.
%
%
We show the stream buffer allocation corresponding to two possible schedules of the inception module.
Both are depth-first schedules, but in Fig.~\ref{fig_dfs_schedule} we start with the leftmost branch, while Fig.~\ref{fig_improved_schedule} starts with the rightmost branch.
This simple change in schedule results in a 30\% reduction in the size of the required stream buffer for this inception module.

When considering large graphs with many branching nodes that either converge to a single output such as GoogLeNet or graphs that diverge to several outputs such as those used for single-shot multibox detection~\cite{ssd}, an exhaustive search of all possible schedules may be infeasible without incurring large compile time penalties.
Our scheduling is conducted using a priority queue based approach, where the cost of executing a given node is determined by the ratio of its output size to its \textit{effective} input size (the size of the input multiplied by the number of users of the input tensor).
This approach allows for the stream buffer savings of Fig.~\ref{fig_improved_schedule} to be achieved, with minimal impact on the compiler runtime.


\comment{
\section{CNN Results}

\subsection{CNN Performance}

\subsubsection{AlexNet, GoogLeNet, SqueezeNet, VGG}
\hl{Author: Mohamed}

\begin{itemize}
  \item Show the highest performance we can get with those 4 networks.
  \item Describe the bottlenecks in each of the networks, and how we overcame them through architecture parameter tuning.
\end{itemize}

\figvs{0.5}{placeholder}{}{Graph showing final performance of these networks -- the text can describe the optimizations performed to get there.}

\subsubsection{ResNet and Resnet-SSD}
\hl{Author: David Han}


\subsection{Accuracy}
\input{accuracy}
}

\section{CNN Implementation}
\label{sec_resnet}


%
This section focuses on 2 popular CNNs: ResNet~\cite{he-corr15} and GoogLeNet~\cite{szegedy2015googlenet}.
We explain different hardware/software co-optimizations that are possible because of our runtime reconfigurable and software programmable overlay.
This allows us to significantly boost the performance of these CNNs on DLA at runtime with little effort, as we show in our results.
%


\subsection{ResNet Convolution Merging}

\begin{figure}[ht]
\centering
\subfloat[Type 1.]{
\includegraphics[width=0.3\linewidth]{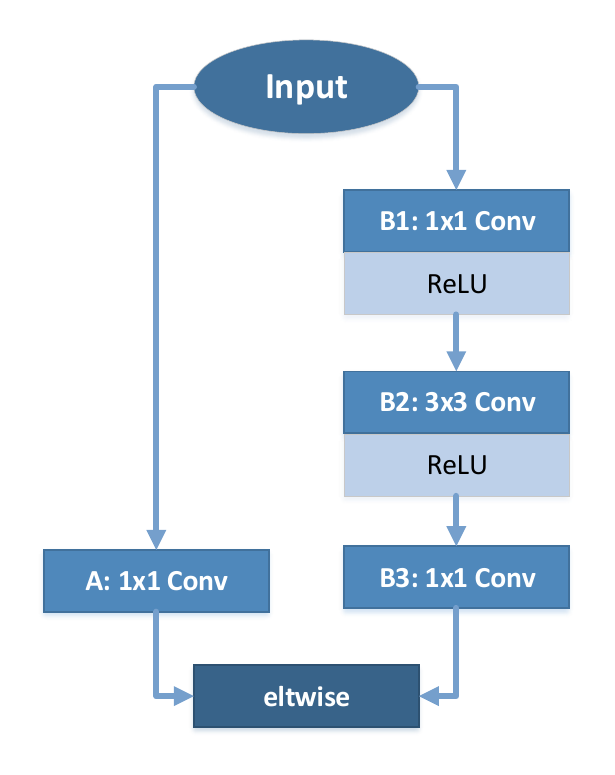}
\label{fig:resmodule-type1}
}
\subfloat[Type 2.]{
\includegraphics[width=0.3\linewidth]{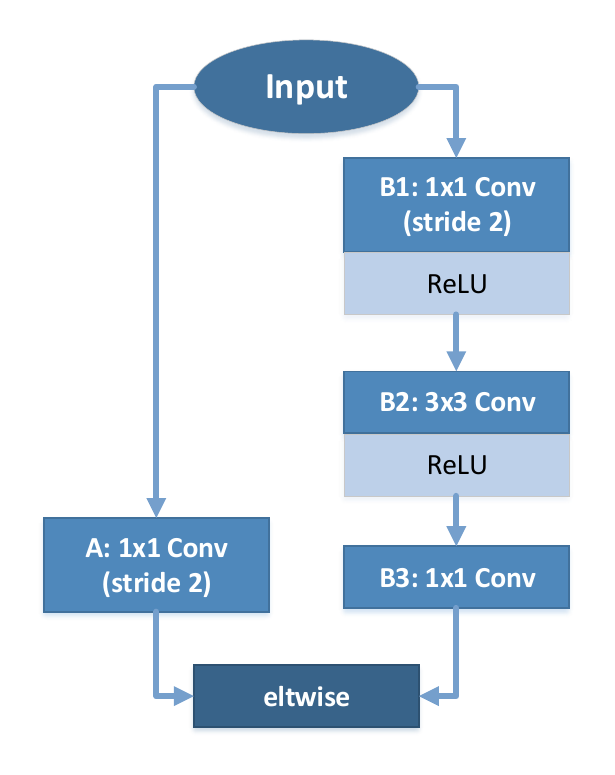}
\label{fig:resmodule-type2}
}
\subfloat[Type 3.]{
\includegraphics[width=0.25\linewidth]{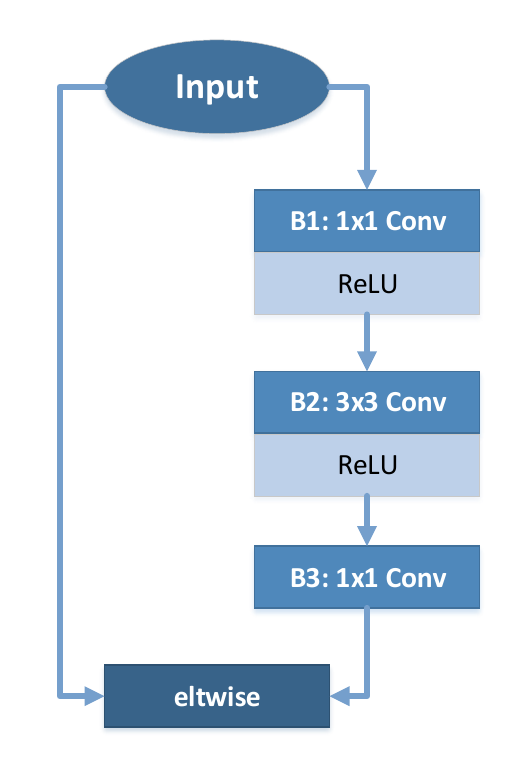}
\label{fig:resmodule-type3}
}
\caption{Types of resmodules in ResNet.}
\label{fig:resmodules}
\end{figure}

ResNet 101 is a large graph that can be targeted for high-definition image resolutions, creating intermediate tensors that require significant slicing to run on the DLA overlay on Arria 10.
ResNet is composed of three types of {\em resmodules}, as shown in Fig.~\ref{fig:resmodules}.
Each type has two convolution branches, merged through an element-wise addition operation ({\em eltwise}).
%


We present a resmodule optimization (implemented automatically in our compiler) that eliminates the eltwise operation by merging it with the preceding convolution(s).
This reduces the total number of arithmetic operations in DLA, and more importantly, decreases the number of slices and DDR4 spill-points.
Instead of storing intermediate tensors between the convolution and the eltwise addition operations, we combine them in a single convolution operation where tensor size is at least half as big as the eltwise input.

\begin{figure}
\centering
\subfloat[Eltwise elimination.] {
\includegraphics[width=0.6\linewidth]{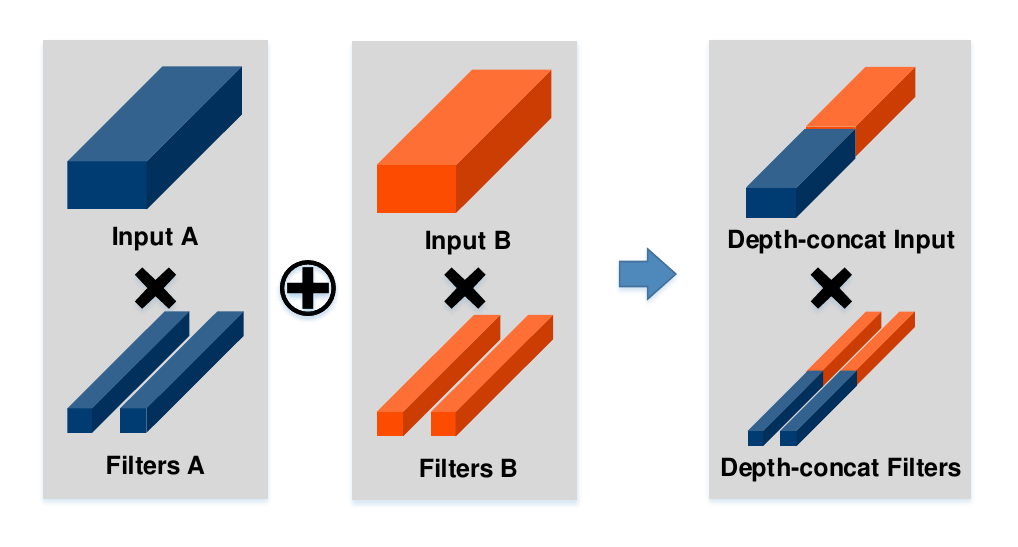}
\label{fig:conv-merging-core}
}
\subfloat[Optimized Type 1 resmodule.] {
\includegraphics[width=0.3\linewidth]{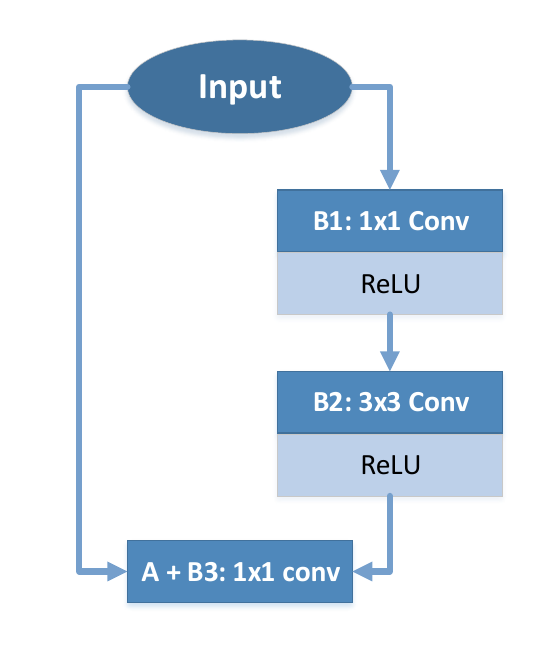}
\label{fig:conv-merging-full}
}
\caption{Convolution merging optimization.}
\label{fig:conv-merging}
\end{figure}

Consider the computation that produces every output element of the eltwise in Type 1 resmodule (Figure~\ref{fig:resmodule-type1}) -- it is the sum of the corresponding output elements of convolution $A$ and $B3$.
As illustrated in Figure~\ref{fig:conv-merging-core}, this sequence of operations is equivalent to a {\em single} convolution after input $A$ and $B3$ (and the corresponding filter $A$ and $B3$) are merged depth-wise.
This effectively absorbs the eltwise addition operation into the dot product operation of the preceding convolutions.
%
%
Figure~\ref{fig:conv-merging-full} shows the Type 1 resmodule after convolution $A$ and $B3$ are merged with the eltwise layer.
Since this optimization converts the explicit eltwise operations into a convolution, output $A$ and $B3$, which would usually reside in DDR4 or on-chip memory, become intermediate results of the merged convolution and are stored in on-chip registers.
This reduction in memory traffic is especially prominent in resmodules, where output $A$ and $B3$ are of 4$\times$ the size of input $A$ and $B3$.


\begin{figure}
\centering
\subfloat[Type 2.] {
\includegraphics[width=0.3\linewidth]{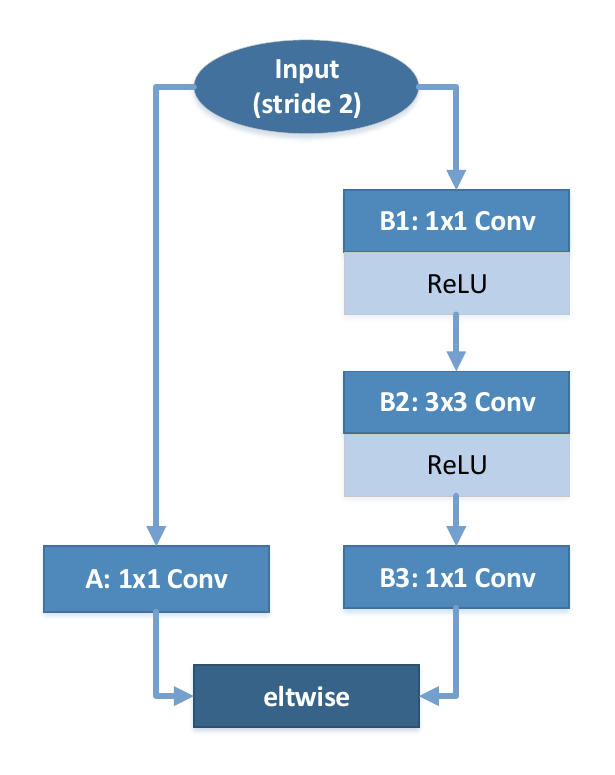}
\label{fig:resmodule-conversion-type2}
}
\quad
\subfloat[Type 3.] {
\includegraphics[width=0.3\linewidth]{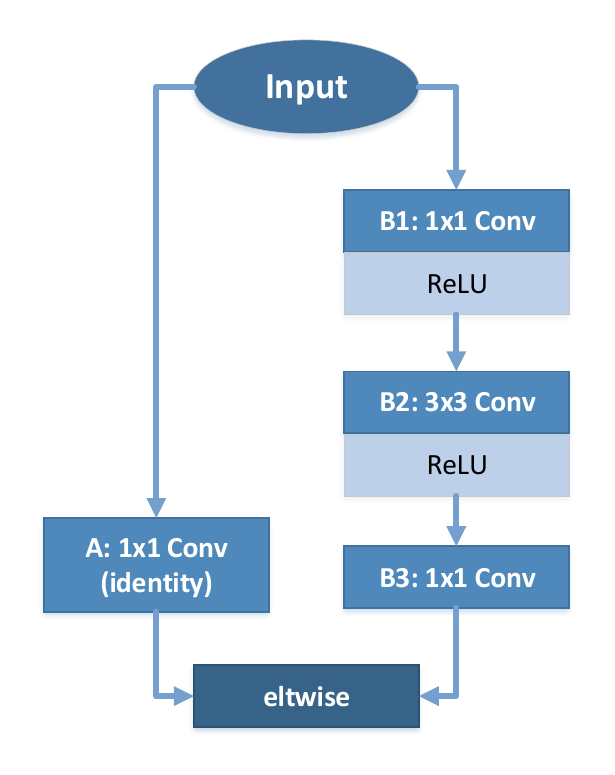}
\label{fig:resmodule-conversion-type3}
}
\caption{Resmodule type conversion to benefit from convolution merging
optimization.}
\label{fig:resmodule-type-conversion}
\end{figure}

In order for Type-2 and Type-3 resmodules to benefit from this optimization, we convert them to Type 1.
For Type 2 (Figure~\ref{fig:resmodule-conversion-type2}), we push the stride-2 convolution $A$ and $B1$ upstream to the layer {\em before} the input.
Not only does this convert the resmodule to Type 1, it also cuts the amount of computation in the upstream layer and reduces the input traffic to convolution $A$ and $B1$.
For Type 3 (Figure~\ref{fig:resmodule-conversion-type3}), we introduce an \textit{identity} convolution -- which creates an identical output tensor from an input tensor -- in the left branch.


\subsection{Non-convolution Primitives}

While almost all layers in ResNet are convolutions, there are a couple of exceptions -- a single Global Average Pooling (GAP) layer and a single Fully-Connected (FC) layer at the end.
This is also true for GoogLeNet where there is a single FC layer, and a single average pooling layer.
Given the extremely low frequency of these non-convolution layers (e.g., 2 out of 147 for ResNet 101), it is best to map them to convolutions.
In this way, we can reuse the powerful convolution engine (PE array) instead of adding dedicated auxiliary kernels that would be under-utilized (over time).


An FC layer performs a multiplication between a vector (input) and a matrix (weights).
It can be mapped to a convolution as follows:
1) the 1D FC input of length $N$ is mapped to a 3D convolution input of shape
   $1 \times 1 \times N$, and
2) the 2D FC weight matrix of shape $N \times M$ is mapped to $M$ 3D
   convolution filters of shape $1 \times 1 \times N$.
With this mapping, the computation of each FC output is assigned to a PE.


Average pooling of window $H \times W$ on a 2D image is equivalent to a 2D convolution with a filter of size $H \times W$.
Each filter element is of value $1 / (H \times W)$.
For a 3D input of depth $D$, average pooling is applied to {\em each} 2D input surface, producing the corresponding output surface.
In this case, the equivalent convolution filter for the output surface at depth $d$, is of shape $H \times W \times D$, with all zero filter values except the surface at depth $d$ being the average pooling filter.
%
%


\subsection{Sparse Filter Shortening}
\label{sec:sparse-filter-shortening}

Even though they save area, the identity and average pooling convolutions introduced in the previous optimizations could come at a high cost to throughput, due to the large but sparse filters involved.
For an identity convolution of input and output shape $H \times W \times D$, there are $D$ filters, each of shape $1 \times 1 \times D$.
Since each filter is responsible for copying input surface at depth $d$ to the output surface at the same depth, the values of this filter are all zeros except 1 at depth $d$.
Fig.~\ref{sparse_shorten} illustrates both the identity and average pooling convolution filters, and how we can leverage their sparsity to conserve operations on DLA.
We improve performance by skipping the computation with filter entries that are filled with zeros.
Since the PEs process $K\_VEC$ filters at a time, we trim the filters size $K\_VEC$ to fit perfectly in the PE array.
This effectively reduces the filter depth from $D$ to $K\_VEC$, saving both compute time and filter data loading time.
We call this optimization {\em sparse filter shortening}, which can also be applied to the average pooling convolution as shown in Fig.~\ref{sparse_shorten}, due to the same filter sparsity.


\figvs{0.9}{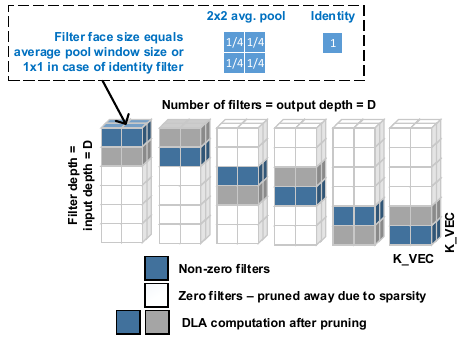}{}{Sparse filter shortening with identity and average-pooling convolution filters.}


\subsection{1x1 Filters Optimization}
\label{sec:filter_packing}


To efficiently compute convolutions using 3x3 filters, the DLA architecture is often tuned to be vectorized in the filter width dimension by setting S\_VEC=3.
Increasing the filter width vectorization increases PE throughput as well as filter prefetch bandwidth for large (eg. 3x3) filters.
However, many of the latest CNNs have a mix of 3x3 and 1x1 filters.
Convolutions using 1x1 filters do not benefit from filter width vectorization, and thus would achieve low DSP efficiency and filter prefetch bandwidth.
To avoid this, the DLA architecture has been optimized for 1x1 filters in two ways.
First, the DSPs that would have been used in a 3x3-filter convolution to process the second and third filter values in the filter width direction are instead used to calculate two additional output pixel in a 1x1-filter convolution.
This allows the PEs to maintain the same DSP efficiency for both 3x3 and 1x1 filters.
Second, the filter prefetch bandwidth added to load a 3-wide filter is used to simply load more 1-wide filters in parallel.
Overall, these two optimizations allow DLA to achieve high throughputs through vectorization for 3x3-filter convolutions without suffering any additional quantization loss for 1x1-filter convolutions.


\subsection{Optimization Impact on ResNet}

Table~\ref{tab:resnet-opt-impact} summarizes the impact of each optimization on the throughput of ResNet 101 with 1080p image resolution.
The number in each row is the normalized throughput after applying {\em all} optimizations listed up to this row.
Here, we apply the mapping of GAP and FC layers to convolution unconditionally (i.e., in the baseline).
The huge speedup of sparse filter shortening comes from the filters of the identity convolutions introduced by convolution merging optimization on Type 3 resmodules which account for 87\% of all resmodules in ResNet 101.


\comment{

\begin{table}[ht]
\centering
\caption{Optimization impact on ResNet-101.}
\label{tab:resnet-opt-impact}
\begin{tabular}{|l|c|c|}
\hline
Optimization  & Relative Throughput & Relative DSP Efficiency \\
\hline
Baseline & 1.0 & 1.0 \\
\hline
Conv. merging (Type 1) & 1.1 & 1.08 \\
\hline
Conv. merging (Type 2) & 1.4 & 1.27 \\
\hline
Conv. merging (Type 3) & 1.7 & 1.00 \\
\hline
Sparse filter shortening  & 3.1 & 0.91 \\
\hline
Group slicing & 3.5 & 1.02 \\
\hline
\end{tabular}
\end{table}
}

\begin{table}[ht]
\centering
\caption{Optimization impact on ResNet-101.}
\label{tab:resnet-opt-impact}
\begin{tabular}{|l|c|}
\hline
Optimization  & Relative Throughput \\
\hline
Baseline & 1.0  \\
\hline
1x1 Filter Opt & 1.3 \\
\hline
Conv. merging (Type 3) & 1.7  \\
\hline
Sparse filter shortening  & 2.8 \\
\hline
Group slicing & 3.1  \\
\hline
\end{tabular}
\end{table}


\subsection{Optimization Impact on GoogLeNet}

Two of the described CNN optimizations are used to improve throughput on GoogleNet: (1) the 1x1 filter optimizations and (2) the average pool mapped to convolution optimization -- this allowed DLA to fit a larger PE array instead of wasting dedicated resources on an average-pooling kernel.
As shown in Table~\ref{tab:goog-opt-impact}, GoogleNet saw a 17\% throughput improvement from these two optimizations.
The following row in the table shows the throughput improvement from increasing the PE array vectorization (from \{P\_VEC,K\_VEC\} = \{1,48\} to \{2,32\}).
Finally, the last row in the table points to an accurate model of external memory optimizations that will allows DLA to achieve \texttildelow900~fps on GoogLeNet on Intel's Arria 10 1150 device, which to our knowledge, is the most efficient acceleration of GoogLeNet on FPGAs.
This optimization entails continuously fetching filters for the next NN layers until the filter cache becomes full instead of limiting filter prefetch only to 1 layer ahead.
While this slightly complicates filter prefetch logic, it has a negligable area cost but allows hiding external memory latency when fetching the NN model.

\begin{table}[ht]
\centering
\caption{Optimization impact on GoogLeNet.}
\label{tab:goog-opt-impact}
\begin{tabular}{|l|c|c|}
\hline
\multirow{ 2}{*}{Optimization}  & Relative & Raw Throughput\\
                                & Throughput & (Intel Arria 10 1150)\\
\hline
Baseline & 1.0 & 469 fps \\
\hline
1x1 Filter Opt & 1.1 & 506 fps \\
\hline
Avg Pool Mapped to Conv & 1.2  & 550 fps \\
\hline
Additional Vectorization & 1.7 & 777 fps \\
\hline
External Memory Opt & 1.9 & \texttildelow900 fps \\
\hline
\end{tabular}
\end{table}


\section{LSTM Cell Implementation}
\label{sec_lstm}


\figvs{1}{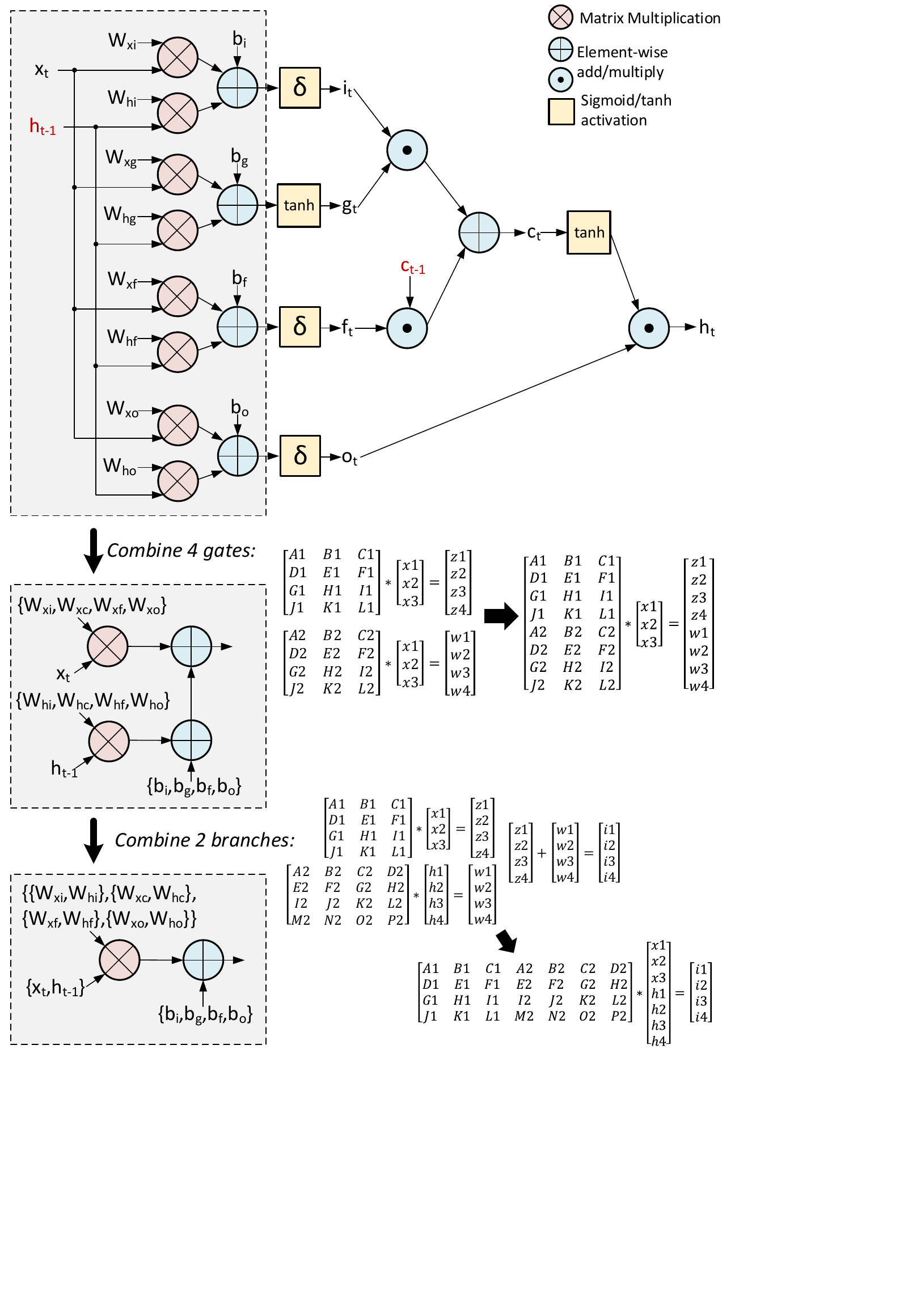}{trim = 0cm 7.5cm 5.5cm 0cm}{Graph/Matrix view of an LSTM cell, and how we combine its matrix-multiplications into one big matrix.}

LSTM cells are a widely-used variant of RNNs, commonly used in speech recognition~\cite{speech_reco}, translation~\cite{wu2016google} and motion detection~\cite{gesture}.
DLA is designed to be a flexible NN accelerator for all relevant deep learning workloads, including LSTM-based networks.
As such, this section discusses how our graph compiler mutates an LSTM cell to map well to the DLA overlay with high performance.

\subsection{Mapping an LSTM Cell to DLA}

Most of the computation in an LSTM cell occurs in 8 matrix multiplications to compute the 3 LSTM gates (input/forget/output)~\cite{ese}.
Fig.~\ref{fig-lstm} illustrates how we combine those 8 matrices into one big matrix -- this reduces DLA execution from 12 subgraphs to a single subgraph which runs at least \texttildelow12$\times$ faster.
First, the 4 matrices that were multiplied by the input/history are each height concatenated as shown in the example in Fig.~\ref{fig-lstm}.
This is a generic optimization that can be applied to any matrix-vector multiplications that share the same input vector.
We end up with two large matrix multiplications, one matrix for the input ($x_t$), and another for the history ($h_{t-1}$).
Next, we combine those two matrices, and the element-wise addition that follows, into one larger matrix through width concatenation of the matrices, and height concatenation of the input and history vectors as shown in Fig~\ref{fig-lstm}.
This gives us one large matrix multiplication for the entire LSTM cell.
Depending on the LSTM cell size, our compiler may later decide to slice this large matrix if it does not fit on the FPGA as described in Section~\ref{sec_slicing}.


\figvs{0.7}{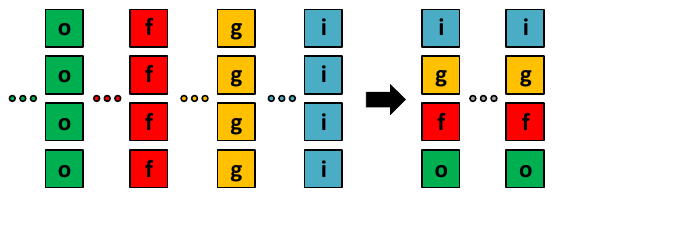}{trim = 0cm 0.8cm 2cm 0cm}{Matrix row interleaving allows streaming different LSTM gate values simulataneously instead of buffering each gate separately.}

\figvs{0.85}{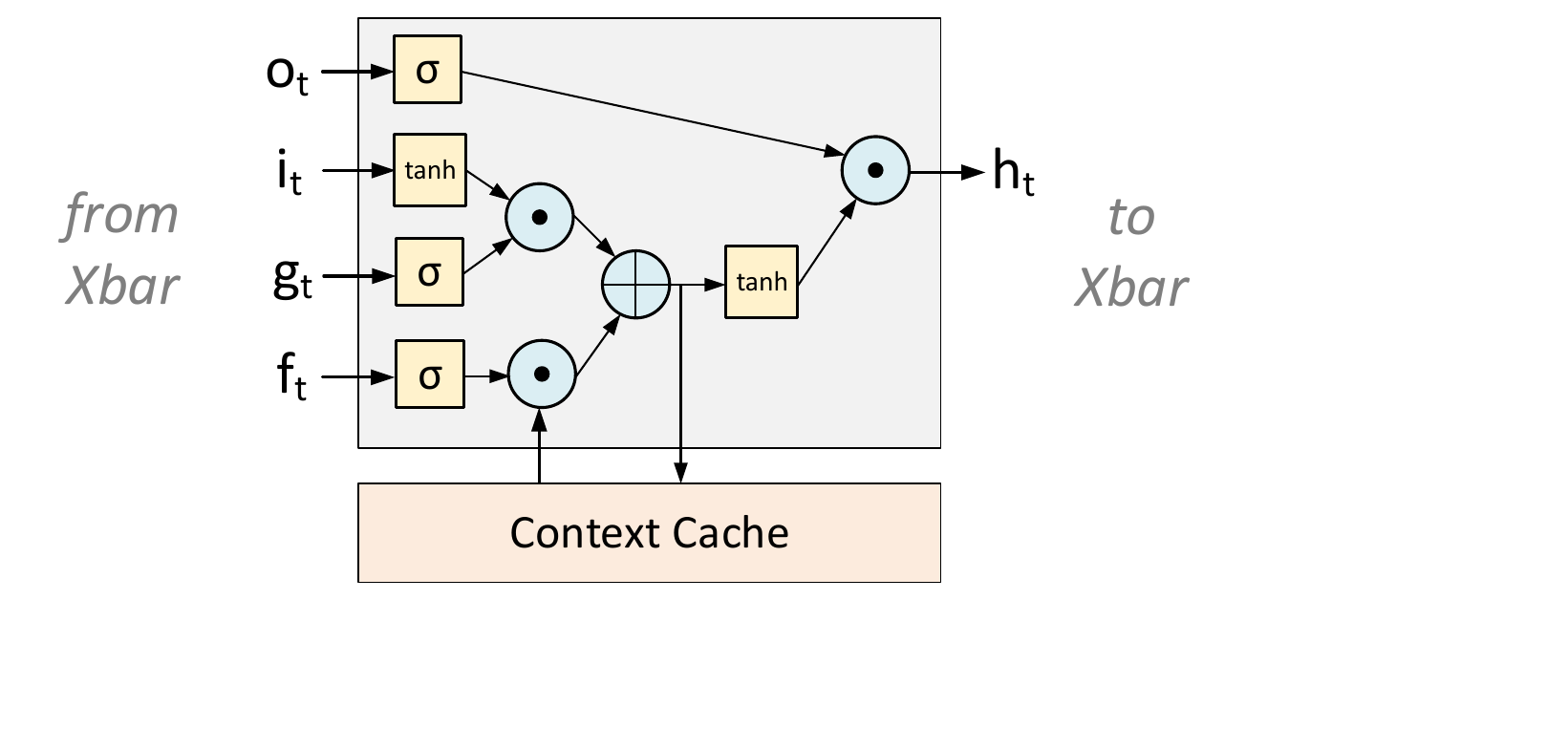}{trim = 0cm 2.7cm 5cm 0cm}{Streaming LSTM hardware block to compute the element-wise operations of an LSTM cell.}

With the combined matrix, each of the LSTM gates are computed one-after-the-other since we compute the matrix rows in order.
However, this is not FPGA-friendly, as each of the input/forget/output gate values will now need to be buffered (using costly on-chip RAM or slow external memory) so that they can be combined in the second half of the LSTM cell.
However, by interleaving the rows of the large matrix (so that the first row contains the filters for the input gate, the second row for the `g' gate, the third row for the forget gate, and the fourth row for the output gate), we can compute one output from each gate in each time step as shown in Fig.~\ref{lstm_interleave}.
This removes the need for buffering large intermediate gate outputs~\cite{cong_lstm}, and allows us to directly stream the gate values into the dedicated LSTM hardware block shown in Fig.~\ref{lstm_block}.

This demonstrates the flexibility of the DLA overlay, and the power of our graph compiler in implementing different NNs.
By simply attaching the LSTM kernel to the Xbar, we can leverage our powerful multi-precision PE array to compute the matrix-multiplication portion of the LSTM cell, then stream data directly into the dedicated LSTM block.


\subsection{External-Memory-bound RNNs}

\figvs{1}{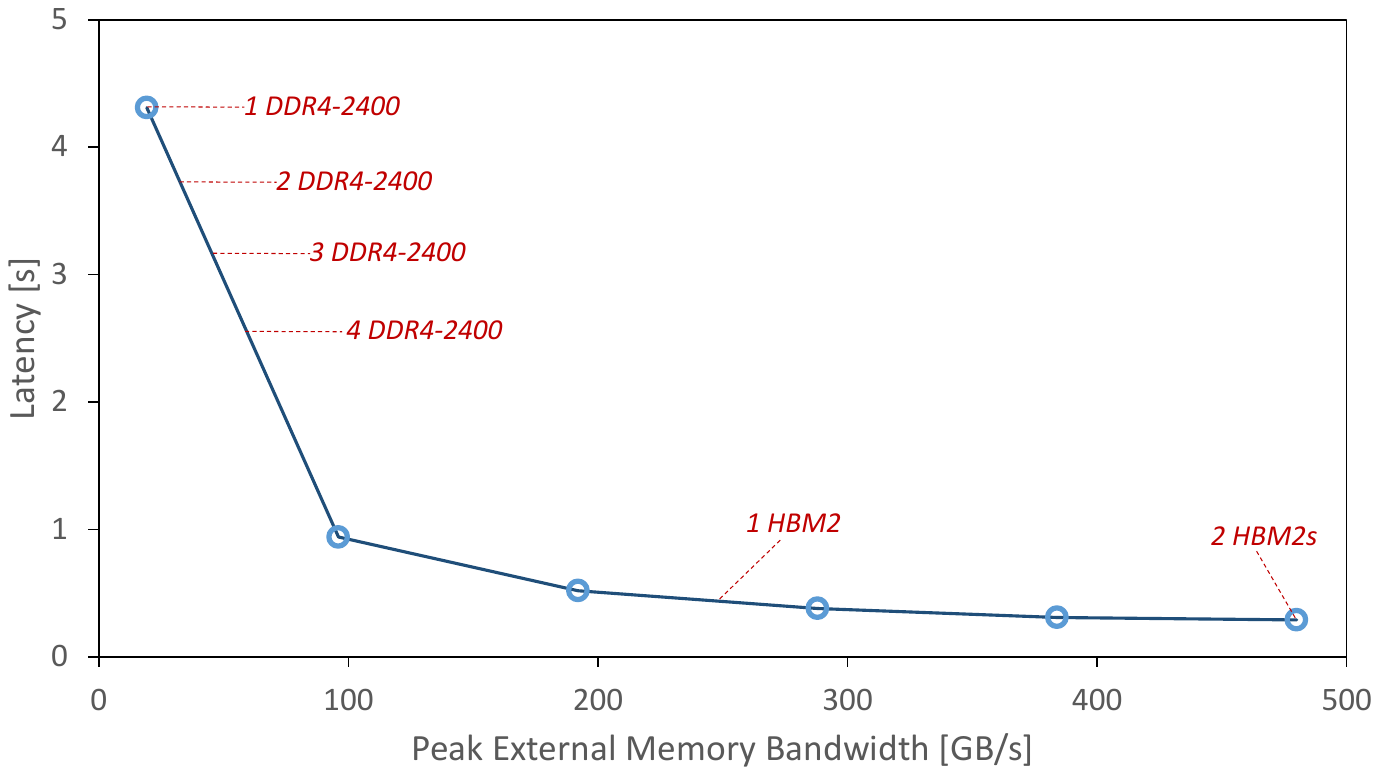}{trim = 1cm 4.5cm 1cm 4.5cm}{Latency of an LSTM NN when varying external memory bandwidth.}

Non-convolutional neural networks, are effectively a matrix-vector multiplication when computed with batch=1.
Most of the applications that use RNNs are real-time applications such as speech/gesture recognition or translation; therefore, they require low-batch and low-latency processing that is ideal for FPGAs.
However, external memory bandwidth is often a bottleneck, since a large matrix has to be fetched from external memory, only to be multiplied with one vector -- compute time is lower than memory fetch time so it is \textit{impossible} to hide memory fetch latency.
Intel's Stratix 10 devices have 2 HBM2 devices integrated on some of their boards, providing up to 500~GB/s of peak memory bandwidth -- this is 20$\times$ higher than a DDR4-2400 memory.
In Fig.~\ref{hbm}, we look towards the future, and model the performance of a 4-layer stacked LSTM NN (with size of input=output=hidden=2048) used for speech recognition.
As the figure shows, with more external memory bandwidth, going from DDR4 to HBM2, the latency for processing a speech segment goes down by more than 5$\times$.


\comment{
\section{Comparison to Other FPGA Solutions}

\comment{
\hl{Author: Andrew Ling}

The idea is to put a big table with all solutions found in the literature and compare them to our numbers. Ideally, everyone can contribute to that table based on the papers they read.
}

Table~\ref{tb_comp} highlights several alternative approaches to CNN acceleration on FPGAs, with the last column illustrating our performance at fp11.  As far as we know, there has never been any previous works that have published in fp11, with no material impact accuracy.  For all the numbers in Table XXX, the authors claim that accuracy is within 1\% of fp32.

\begin{table*}[!t]
\centering
\begin{small}
    \caption{Comparison of recently-published FPGA NN accelerators.}
    \label{tb_comp}
    \begin{tabular}{|l|l|l|l|l|l|}
      \hline
	                      & ICCAD'16     & FPGA'17         & FPGA'17        & FPGA'17         & DLA  \\\hline
	  Design Input        & Xilinx HLS   & RTL             &  OpenCL        & OpenCL + SV     & OpenCL  \\\hline
	  Precision           & Fixed 16-bit & Fixed 8/16 bits & FP16           & Fixed 8/16 bits & FP16/11/8  \\\hline
	  Platform            & Kintex KU060 & Arria 10 1150   & Arria 10 1150  & Arria 10 1150   & Arria 10 1150  \\\hline
	  Model (GOPs)        & VGG (30.76)	 & VGG (30.76)	   & AlexNet (1.45)	& VGG (30.76)	    & GoogLeNet (3.2)  \\\hline
	  Throughput (GOPs/s) & 266          & 645.25          &  1382          & 1790            & 1760  \\\hline
	  Efficiency (\%)     & 62.8\%       &	68.6\%	       & 91\%	          & 84\%	          & 90\%  \\\hline
    \end{tabular}
\end{small}
\end{table*}
%
%


\section{The Memory Wall}
\comment{
\hl{Author: Andrew Ling}

\begin{itemize}
  \item Look-ahead to s10 and show scaling?
  \item Multi-FPGA solution and how we break the memory wall. \hl{Author: Andrew Bitar}
  \item HBM advantage -- show LSTM modeling?
  \item Other detailed implementation things that are in-progress.
\end{itemize}

\hl{M: thinking of changing this section name to ``the memory wall", and focus on HBM and multi-fpga exclusively.}
\hl{M: Say something like: we can perfect the compute on-chip by going to lower-precision, and improve vectorization, and do tricks like WB, sparsity, etc. But you will always face memory problems -- hence our future focus on HBM/multi-fpga}

As shown in the previous sections, one of the key advantages of the FPGA is its low-latency on-die memory block rams.
Unlike other devices which must store large amounts of data externally, the FPGA can keep the data on-die close to the core compute leading to high PE and energy efficiency.
In our current work, however, we still must store the model external since the capacity of model parameters exceeds the capacity of the on-die storage.
In future work, we hope to get around this limitation by either compressing the model through quantization techniques, similar to what has been explored previously in [XXX,XXX].
Additionally, by splitting the model across multiple FPGAs, we can also expand the on-die storage capacity, without having to increase the number of model parameters to store.  Both of these techniques may provide enough on-die storage such that the entire model can live directly on the FPGA, without having to use any external memory, further boosting the overall performance and energy efficiency of the FPGA.
Additionally, current devices, such as Intel’s Stratix 10, have significantly more on-die memory in a single device, allowing us to cache the entire model on a single Stratix 10 device.
In addition to increased on-die capacity, Stratix 10 also provides access to HBM interfaces, which significantly boosts the external memory bandwidth.
For some critical low-latency applications, this can be taken advantage of by allowing for more complicated models to be processed on the FPGA.
An example of how we can take advantage of this is shown in the following figure, which shows the impact of LSTM through with respect to external memory bandwidth.
As the figure shows, for large external memory bandwidth requirements of 200GB/s or more, latency improves. Such a high external bandwidth requirements is generally best served with HBM.
}

The two cornerstones of FPGA acceleration are compute efficiency and memory bandwidth -- balancing the two is key to high performance.
With Minifloat, we can double or triple our performance because we can fit more operations on the chip.
However, external memory bandwidth quickly becomes a performance bottleneck, especially with ever-larger CNN models like ResNet-101, and RNNs that have far less filter reuse and also large models.
Looking towards the future, we present two ways of breaking this ``memory wall" to overcome external memory bandwidth limitations and continue improving performance.

\subsection{High Bandwidth Memory (HBM)}

\figvs{1}{hbm}{trim = 1cm 5cm 1cm 5cm}{Latency of an LSTM NN when varying external memory bandwidth.}

Non-CNN neural networks, such as LSTMs, are effectively a matrix-vector multiplication when computed with batch=1.
Most of the applications that use RNNs are real-time applications such as speech/gesture recognition or translation; therefore, they require low-batch and low-latency processing that is ideal for FPGAs.
However, external memory bandwidth is often a bottleneck, since a large matrix has to be fetched from external memory, only to be multiplied with one vector -- compute time is lower than memory fetch time so it is \textit{impossible} to hide memory fetch latency.

Intel's Stratix 10 devices have 2 HBM2 devices integrated on some of their boards, providing up to 500~GB/s of peak memory bandwidth -- this is 20$\times$ higher than a DDR4-2400 memory.
In Fig.~\ref{hbm}, we model the performance of a 4-layer stacked LSTM NN used for speech recognition.
As the figure shows, with more external memory bandwidth, going from DDR4 to HBM2, the latency for processing a speech segment goes down by more than 5$\times$.


\subsection{Multi-FPGA Systems}

\hl{cite microsoft}
\comment{
\begin{itemize}
  \item Off-chip memory BW can bottleneck deeper graphs like ResNet101
  \item This would be fixed if we had more on-chip memory
  \item Can get more on-chip memory by using more FPGAs, connect them through high BW tranceivers
\end{itemize}
}

One of the key advantages of the FPGA is its low-latency on-die memory block RAMs.
Unlike other devices which must store large amounts of data externally, the FPGA can keep the data on-die close to the core compute leading to high PE and energy efficiency.
However, the ever increasing memory requirements needed to accelerate the latest neural networks, such as ResNet101, has forced FPGA accelerators to use a mix of on-chip and off-chip memory (see Section~\ref{sec:mem-arch}).
Although the DLA Software Compiler can perform graph optimizations to mitigate the impact of spilling to off-chip memory (Section~\ref{sec:slicing}), the achievable accelerator performance will still be bottlenecked by DDR bandwidth.
Other than some high external memory bandwidth platforms (as described in Section~\ref{sec:high-bw-mem}), most FPGA platforms have limited external memory bandwidth compared to other devices.
Thus, most FPGA accelerators are particularly interested in avoiding off-chip memory.

DLA provides a means to avoid off-chip memory in larger neural networks by using multiple FPGA devices, as is illustrated in Figure~\ref{fig:multi-node}.
By leveraging the high bandwidth transceivers to send filter and feature data between multiple FPGAs, DLA can distribute large neural network models across the block RAMs of multiple FPGAs and thus avoid spilling to DDR.
Once a certain application has enough devices to provide enough on-chip memory to avoid external memory, we describe the application as surpassed ``The Memory Wall''.

This Memory Wall can be clearly visualized in Figure~\ref{fig:memory-wall}, where the performance per FPGA of a ResNet-101 network is plotted against number of Arria 10-1150 devices.
As can be seen in the figure, once there are enough FPGAs to fit the entire network on-chip, the performance per device sees a jump.
Before and after the Memory Wall, additional FPGAs see diminishing returns on performance due to quantization effects.
Once there are enough devices to avoid the bottlenecks from external memory bandwidth, the performance is seen to overcome the Memory Wall.

\begin{figure}[!t]
\centering
\includegraphics[width=\columnwidth,keepaspectratio]{images/multi-node-crop}
\caption[Multi-Node]{DLA implemented on multiple FPGAs.}
\label{fig:multi-node}
\end{figure}

\begin{figure}[!t]
\centering
\includegraphics[width=\columnwidth,keepaspectratio]{images/memory-wall-crop}
\caption[Performance vs Num FPGAs]{ResNet-101 performance per FPGA with increasing number of A10-1150 devices.}
\label{fig:memory-wall}
\end{figure}

\comment{
\subsection{Stratix 10 and On-Chip Memory}

Intel's latest generation of FPGA -- the Stratix 10 device -- provides additional on-chip memory on a single FPGA, reducing the number of devices needed to overcome the Memory Wall.
Particularly for the classification of SD images, the S10 device can avoid using external memory while accelerating a large network on a single FPGA.
Table~\ref{tbl:s10} includes throughput projections based on DLA performance models.
Particularly at low-batch, the S10 device will achieve highly competitive single-device performance.
Additionally, fewer S10 devices will be needed to overcome the Memory Wall for neural network acceleration of HD images, providing a highly-efficient solution in terms of both power and cost.

\begin{table}[!t]
\centering
\begin{small}
    \caption{Throughput projections on S10-2800 for SD images using FP11.}
    \label{tbl:s10}
    \begin{tabular}{|l|c|c|}
      \hline
          Graph               & Batch=1 Throughput   \\\hline
          GoogleNet           & 3,930 imgs/s      \\\hline
          SqueezeNet          & 8,500 imgs/s         \\\hline
          ResNet-18           & 4,440 imgs/s       \\\hline
          ResNet-50           & 1,290 imgs/s       \\\hline
          ResNet-101          & 710 imgs/s         \\\hline
    \end{tabular}
\end{small}
\end{table}
}

\comment{
\begin{itemize}
  \item S10 provides lots of on-chip memory
  \item Makes it easier to break the memory wall for SD images
  \item Models show we will achieve these throughputs on S10
  \item HD images could leverage multiple S10 devices to break the memory wall
\end{itemize}
}


}

\section{Conclusion}

We presented a methodology to achieve software ease-of-use with hardware efficiency by implementing a domain specific customizable overlay architecture.
We described the hardware tradeoffs involved with NN acceleration, and delved into our graph compiler that maps NNs to our overlay.
We then showed that, using both our hardware and software, we can achieve 3$\times$ improvement on ResNet 101 HD, 12$\times$ on LSTM cells, and 900~fps on GoogLeNet on Intel's Arria~10 FPGAs.
We will further develop DLA to encompass more use-cases such as multi-FPGA deployment~\cite{microsoft_brainwave}.
In the future, we also aim to implement similar overlays for different application domains such as genomics, packet processing, compression and encryption to further make FPGAs accessible for high-throughput computation.


\bibliographystyle{abbrv}
\bibliography{dla18}

\begin{thebibliography}{10}

\bibitem{previous_work}
U.~e.~a. Aydonat.
\newblock An opencl\texttrademark deep learning accelerator on arria 10.
\newblock In {\em FPGA}, FPGA '17, pages 55--64, New York, NY, USA, 2017. ACM.

\bibitem{microsoft_brainwave}
E.~e.~a. Chung.
\newblock Accelerating persistent neural networks at datacenter scale.
\newblock HotChips, 2017.

\bibitem{speech_reco}
A.~G. et~al.
\newblock Speech recognition with deep recurrent neural networks.
\newblock {\em ICASSP}, 2013.

\bibitem{bad_overlay}
A.~J. et~al.
\newblock Efficient overlay architecture based on dsp blocks.
\newblock {\em FCCM}, 2015.

\bibitem{krizhevsky2012alexnet}
A.~K. et~al.
\newblock Imagenet classification with deep convolutional neural networks.
\newblock In F.~Pereira, C.~J.~C. Burges, L.~Bottou, and K.~Q. Weinberger,
  editors, {\em Advances in Neural Information Processing Systems 25}, pages
  1097--1105. Curran Associates, Inc., 2012.

\bibitem{szegedy2015googlenet}
C.~S. et~al.
\newblock Going deeper with convolutions.
\newblock In {\em Computer Vision and Pattern Recognition (CVPR)}, 2015.

\bibitem{cnn1}
H.~Z. et~al.
\newblock A framework for generating high throughput cnn implementations on
  fpgas.
\newblock {\em FPGA}, 2018.

\bibitem{cnn2}
J.~S. et~al.
\newblock Towards a uniform template-based architecture for accelerating 2d and
  3d cnns on fpga.
\newblock {\em FPGA}, 2018.

\bibitem{opencl}
M.~A. et~al.
\newblock Gzip on a chip: High performance lossless data compression on fpgas
  using opencl.
\newblock {\em IWOCL}, 2014.

\bibitem{swish}
P.~R. et~al.
\newblock Swish: a self-gated activation function.
\newblock {\em AISTATS}, 2017.

\bibitem{ese}
S.~H. et~al.
\newblock Ese: Efficient speech recognition engine with sparse lstm on fpga.
\newblock {\em FPGA}, 2017.

\bibitem{ssd}
W.~L. et~al.
\newblock Ssd: Single shot multibox detector.
\newblock {\em ECCV}, 2016.

\bibitem{cong_lstm}
Y.~G. et~al.
\newblock Fpga-based accelerator for long short-term memory recurrent neural
  networks.
\newblock {\em ASP-DAC}, 2017.

\bibitem{he-corr15}
K.~He, X.~Zhang, S.~Ren, and J.~Sun.
\newblock Deep residual learning for image recognition.
\newblock {\em CoRR}, abs/1512.03385, 2015.

\bibitem{tensorflow}
G.~Inc.
\newblock Tensorflow, 2018.

\bibitem{gesture}
K.~Murakami and H.~Tagushi.
\newblock Gesture recognition using recurrent neural networks.
\newblock {\em SIGCHI}, 1991.

\bibitem{simonyan2014very}
K.~Simonyan and A.~Zisserman.
\newblock Very deep convolutional networks for large-scale image recognition.
\newblock {\em arXiv preprint arXiv:1409.1556}, 2014.

\bibitem{wu2016google}
Y.~e.~a. Wu.
\newblock Google's neural machine translation system: Bridging the gap between
  human and machine translation.
\newblock {\em arXiv preprint arXiv:1609.08144}, 2016.

\end{thebibliography}

\end{document}